\documentclass[aps,twocolumn,showpacs]{revtex4-1}
\usepackage{float}
\usepackage{makeidx}
\usepackage{amsmath,amssymb,graphics,epsfig,color,times,bbm}

\begin{document}

\title{Experimental realization of three-color entanglement at optical fiber communication and atomic storage wavelengths}
\author{Xiaojun Jia, Zhihui Yan, Zhiyuan Duan, Xiaolong Su, Hai Wang, Changde Xie}
\author{Kunchi Peng}
\email{kcpeng@sxu.edu.cn}

\affiliation{State Key Laboratory of Quantum Optics and Quantum Optics Devices,\\
Institute of Opto-Electronics, Shanxi University, Taiyuan, 030006,
People's Republic of China}

\begin{abstract}
Entangled states of light including low-loss optical
fiber transmission and atomic resonance frequencies are essential
resources for future quantum information network. We present the
experimental achievement on the three-color entanglement generation
at $852$ $nm$, $1550$ $nm$ and $1440$ $nm$ wavelengths for optical continuous
variables. The entanglement generation system consists of two
cascaded non-degenerated optical parametric oscillators (NOPOs). The
flexible selectivity of nonlinear crystals in the two NOPOs and the
tunable property of NOPO provide large freedom for the frequency
selection of three entangled optical beams. The presented system is
hopeful to be developed as a practical entangled source used in quantum
information networks with atomic storage units and long fiber
transmission lines.
\end{abstract}


\maketitle

Entanglement is the most typical quantum feature that has no analogue in classical physics. In the development of modern physics the understanding to the conception and property of entanglement has continually attracted the study interests of both theoretical and experimental physicists \cite{Hor,Brau,Reid}. Especially, it has been demonstrated that quantum entanglement is the most important resource in quantum communication and computation. A lot of endeavor has been paid in preparing various quantum entangled states over past twenty years \cite{Bou,Ou,Fur,Li,Jia2,Jing2,Yon,Lance}. A variety of bipartite optical continuous variable (CV) entangled states have been generated and applied in different protocols of quantum communication involving two parties \cite{Fur,Li,Jia2}. However, a real quantum information network should be composed of many nodes and channels \cite{Jing2,Yon,Lance}. It has been known that the controlled quantum communications only can be achieved under the help of multipartite (more than two parties) entangled states. Based on the use of tripartite CV entangled states the interesting quantum communication experiments, such as controlled dense-coding \cite{Jing2}, quantum teleportation network \cite{Yon} and quantum secret sharing \cite{Lance} etc., have been achieved.  Toward practical applications in the real-world we have to establish quantum information networks (QINs) involving both light
and matter atoms, where light is used for communicating among
distant nodes consisting of matter atoms \cite{Kimble}. The atomic systems
in network nodes serve as the storages of quantum information. The storage and retrieve of quantum states of light are the important operations realizing QINs and have been experimentally demonstrated based on Cs and Rb atoms by several groups \cite{Appel,Honda,Cvi,Jensen,Wang1}. For developing practical CV
QIN with both nodes and fiber transmission lines, it is
essential to prepare multi-partite entangled states consisting of optical sub-modes at fiber transmission and atomic transition frequencies.

    Non-degenerate optical parametric oscillators (NOPOs) above the
threshold are the most successful devices for producing two-color and multi-color
CV entangled optical beams in the achieved experiments of quantum
optics \cite{Villar,Su,Jing,Kell,Guo,Coe}.  Two-color entangled optical beams
at different frequency regions have been experimentally prepared by means of above-threshold NOPOs with various pump lasers and nonlinear crystals \cite{Villar,Su,Jing,Kell,Guo}. Recent years, for satisfying the requirements of the developing QIN the generation schemes of multi-color CV entangled states via intra-cavity nonlinear processes have been theoretically proposed  \cite{Vill,Cass,Tan,Gu}. In 2009, the first CV three-color entangled state was experimentally produced by an above-threshold NOPO at a low temperature of $-23$ $^0C$ \cite{Coe}. The three entangled sub-modes in this experiment are the output signal, idler and reflected pump modes from a NOPO and their wavelengths are $1062.102$ $nm$, $1066.915$ $nm$ and $532.251$ $nm$, respectively. It has been theoretically and experimentally demonstrated that the spurious excess phase noise introduced by the non-linear crystal into the intra-cavity pump field \cite{Cas,Ces}, which derives from the influence of phonon noise, will inevitably destroy the phase correlations among the sub-modes. For reducing the influence of the phonon noise on the reflected pump field a new scheme of producing three-color entanglement without the use of the reflected pump field is desired. The discrete-variable (DV) entangled state of triple photons at three different wavelengths are experimentally produced by a cascaded spontaneous parametric  down-conversion process in two different nonlinear crystals \cite{Hub,Shalm}. Transferring the scheme producing three-color DV entangled photons to CV regime, our group proposed a generation system of CV three-color entangled optical beams, in which two cascaded NOPO1 and NOPO2 are utilized  \cite{Tan}. One of the entangled signal and idler optical beams produced by NOPO1, is used for the pump light of  NOPO2. The three-color entanglement among signal and idler beams produced by NOPO2 and the retained another output beam from NOPO1 is theoretically demonstrated and the optimal operation conditions of the cascaded NOPOs system are numerically calculated in Ref. \cite{Tan}. Following the theoretical design we have achieved the experimental generation of three-color CV entangled state by using the cascaded NOPOs system for the first time. Through the special selections of the pump laser and the nonlinear crystals in the two  NOPOs, the wavelength of one of the obtained three-color entangled beams is $852$ $nm$ which can be tuned to a transition frequency of Cs atoms thus can be used for the storage of quantum information. The wavelengths of other two beams are $1550$ $nm$ matched for optimal transmission in optical fibers and $1440$ $nm$ close to fiber window with quite low (although not optimal) transmission losses \cite{Pal}. The produced three-color CV entangled states are suitable to be applied in the future quantum information networks containing both atomic storage unit and optical fiber transmission line.

\begin{figure}
\centerline{
\includegraphics[width=86mm]{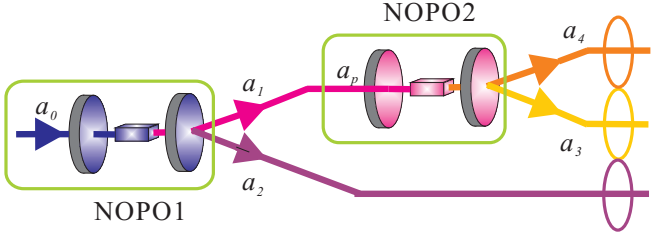}
} \caption{(Color online) The principle schematic of the three-color entanglement
generation system.  \label{Fig1} }
\vspace{-1em}
\end{figure}

Fig.1 is the principle schematic of the three-color entanglement generation system, which consists of NOPO1 and NOPO2. The NOPO1 is pumped by a laser ($a_{0}$) to create a pair of intense optical beams ($a_{1}$ and $a_{2}$), the
frequencies of which fulfill energy conservation $\omega _{0}=\omega _{1}+\omega _{2}$,
where the subscripts $j=0,1,2$ designate the pump
beam and the two generated down-conversion optical beams, respectively. One of the generated
optical beams ($a_{p}^{in}=a_{1}^{out}$) drives the NOPO2 to achieve the cascaded intra-cavity optical parametric down-conversion and to produce the output fields $a_{3}$ ($\omega _{3}$) and $a_{4}$ ($\omega _{4}$), the frequency sum of which equals to $\omega _{1}$ ($\omega _{1}=\omega _{3}+\omega _{4}$).
It has been theoretically proved in Ref. \cite{Tan} that the three final output
light beams ($a_{2}$, $a_{3}$ and $a_{4}$) originating from a single pump
beam ($a_{0}$) have strong intensity correlation. On the other hand,
the frequency constraint among the three optical modes translates into a constraint for the phase
variations, so the phase fluctuations of the optical modes $a_{1}$ and $a_{2}$ ($a_{3}$ and $a_{4}$) should be anti-correlated each other, the sum of their phase fluctuations should be correlated with the phase fluctuation of the pump field. The experimental criteria in terms of the variances of particular
combinations of the amplitude and phase quadratures for testing genuine CV multipartite entanglement among optical modes have been given by Loock and Furusawa \cite{Loock}. A set of the criterion inequalities for determining the CV tripartite entanglement is expressed by:

\begin{eqnarray}
\Delta _{1} &=&\langle \delta ^{2}(X_{3}-X_{4})\rangle +\langle \delta
^{2}(g_{1}Y_{2}+Y_{3}+Y_{4})\rangle \geq 4,  \notag \\
\Delta _{2} &=&\langle \delta ^{2}(X_{2}-X_{4})\rangle +\langle \delta
^{2}(Y_{2}+g_{2}Y_{3}+Y_{4})\rangle \geq 4,  \notag \\
\Delta _{3} &=&\langle \delta ^{2}(X_{2}-X_{3})\rangle +\langle \delta
^{2}(Y_{2}+Y_{3}+g_{3}Y_{4})\rangle \geq 4.
\end{eqnarray}
Where, ''4'' is the boundary to verify the full inseparability of the
genuine tripartite entanglement. The $X_{j}$ and $Y_{j}$ ($j=2$, $3$, $4$)
represent the amplitude quadratures and the phase quadratures of the
resultant output modes $a_{2}$, $a_{3}$ and $a_{4}$, respectively. Each
pairs of $X_{j}$ and $Y_{j}$ satisfy the canonical commutation relation $%
[X_{j},Y_{j}]=2i$. The $g_{j}$ ($j=1$, $2$, $3$) are the parameter gain factors (arbitrary real parameters), which are chosen to minimize the combined correlation variances at the left side of the inequalities. The theoretical analysis has
demonstrated that if any two inequalities in the set of inequalities (1) are
simultaneously violated the three optical modes $a_{2}$, $a_{3}$ and $a_{4}$
are in a CV entangled state \cite{Loock}. In the experiments we should select the optimal gain factors ($g_{j}^{opt}$) to obtain the minimal correlation variance combinations, which can be accomplished by adjusting the electronic gains of the detected photocurrents as described in Supplemental Material \cite{SM}).

\begin{figure}
\centerline{
\includegraphics[width=86mm]{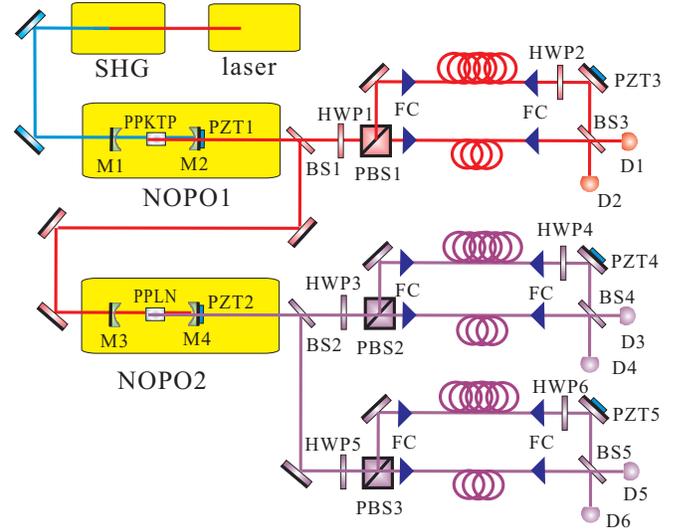}
} \caption{(Color online) Schematic of the experimental setup. Laser: Ti: Sapphire laser; NOPO$_{1-2}$: non-degenerate optical parametric oscillator; M$_{1-4}$: different mirror; BS$_{1-5}$: different beam splitter (see context for detailed); HWP: half wave plate; FC: fiber coupler;
PZT$_{1-5}$: piezoelectric transducer; D$_{1-6}$: high efficiency detector for different wavelength. \label{Fig2} }
\vspace{-1em}
\end{figure}

The experimental setup is shown in Fig. 2. The coherent optical field at $795$ $nm$ from a continuous-wave Ti: Sapphire laser (MBR110, Coherent Ltd.) is frequency-doubled by a
second harmonic generator (SHG) to obtain the light at $398$ $nm$. Both NOPO1 and NOPO2 are in a Fabre-Perot cavity configuration consisting of two concave mirrors with a 50-mm radius curvature (M1 and M2 for NOPO1, M3 and M4 for NOPO2). For generating the multi-color entangled state satisfying different frequency requirements the wavelength difference between the signal and the idler optical beams produced by NOPOs is large usually. The large wavelength difference must result in a large walk-off effect between the two beams and thus significantly decrease the effective interaction length in the non-linear crystal \cite{Boyd}. The periodically poled (PP) crystals have the structure of regularly spaced ferroelectric domains with alternating orientation, which can effectively overcome the harmful walk-off effect \cite{Miss}. The nonlinear coefficient of PPLN (PP lithium niobate) is higher than that of PPKTP (PP Potassium Titanyl Phosphate), but its absorption loss is larger than that of PPKTP  \cite{Miss,Eger}. Since the idler output beam of NOPO1 is used for the pump light of NOPO2, its intensity should be higher than the threshold pump power of NOPO2 at least. For producing intense output from NOPO1 the PPKTP with lower absorption loss is chosen for reducing its intra-cavity loss. To NOPO2, we hope that it has the higher parametric conversion efficiency and lower threshold pump power so the PPLN with the higher nonlinear coefficient is utilized. The size of both PPKTP and PPLN is $1*2*10$ $mm^3$. The two nonlinear crystals are respectively placed inside an oven, the temperature of which can be independently well-controlled and tuned by an electronic temperature-controller with the precision of $0.01$ $^0C$ (YG-IIS-RA, Yuguang Ltd.). M1 (M3) is used for the input coupler of NOPO1 (NOPO2) with the transmissivity of $30\%$ ($10\%$) at $398$ $nm$ ($746$ $nm$) and the high reflectivity at $746$ $nm$ and $852$ $nm$ ($1440$ $nm$ and $1550$ $nm$). M2 (M4) is utilized as the output coupler with the high reflectivity at $398$ $nm$ ($746$ $nm$) and the transmissivity of $3.0\%$ ($4.0\%$) at $746$ $nm$ and $852$ $nm$ ($1440$ $nm$ and $1550$ $nm$). M2 (M4) is mounted on piezoelectric transducer (PZT1 (PZT2)) for scanning actively the cavity length of NOPO1 (NOPO2) or locking it on the resonance with the generated sub-harmonic modes as needed. The cavity length, the finesse and the threshold pump power for NOPO1 (NOPO2) are $101.5$ $mm$ ($101.9$ $mm$), $195$ for $746$ $nm$ ($149$ for $1550$ $nm$) and $75$ $mW$ ($4.5$ $mW$), respectively.

Three M-Z interferometers with unbalanced arm lengths are applied to measure
the noise powers of the phase and the amplitude quadratures for the three
resultant subharmonic modes at $852$ $nm$ (generate by NOPO1), $1440$ $nm$
and $1550$ $nm$ (generated by NOPO2) as well as to determine the
corresponding quantum noise limits (QNLs) \cite{Su,Glo}. The three
interferometers have the identical configuration, each of which consists of
a polarizing-beam-splitter (PBS1-3), two high reflection mirrors and a 50/50
beam-splitter (BS3-5). A PZT (PZT3-5) is mounted on a reflection mirror of
the interferometer for locking the relative phase between the long and the
short arms of the unbalanced interferometer to the required value. The
half-wave plate (HWP1-6) is used for aligning the polarization direction of
the optical mode. Two polarization-preserved optical fibers with different
lengths, in each end-face of which a fiber coupler (FC) is attached, serve as the long and
the short arms of the interferometer, respectively. The noise powers of the
output optical beams from the interferometers are detected by the
photo-diodes (D1-D6). For measuring the correlation variance among
the three entangled sub-modes ($a_{2}$, $a_{3}$ and $a_{4}$), we firstly
detect the photocurrent noise power of the amplitude quadrature or the phase
quadrature of each sub-mode and then combine them with the positive or the
negative power combiners according to the requirement of Eqs. (1). It has
been well-proved in Refs. \cite{Su,Glo,Wang}, when the input optical beams
only passes through the short arm of the interferometer, which can be
completed by aligning the polarization of the input light to the
transmission direction of PBS1-3 with the HWP in front of the PBS, the sum
and the difference of the photo-currents detected by a pair of D1 and D2,
(also D3 and D4, D5 and D6) are the noise power of the amplitude quadrature
of the input beam and the corresponding QNL, respectively. If keeping the
optical phase difference between the long and the short arms at $\pi
/2+2k\pi $ ($k$ is an integer) and splitting equally the input optical beam
in the two arms of the interferometer, which can be realized by aligning
the polarization of the input light to $45^{o}$ of PBS1-3 polarization orientation by rotating the
HWP, the difference (sum) of the photo-currents detected by a pair of the
detectors is the noise power of the phase quadrature of the input beam
(corresponding QNL). On the other hand, for implementing the noise power
measurement of an input optical beam at a specific noise sideband ($f$), the
phase shift ($\theta $) of the spectral component at $\Omega =2\pi f$
between the two arms of the M-Z interferometer should be controlled at $%
\theta =n\Omega \Delta L/c=\pi $ ($\Delta L$ is the length difference
between the two arms; $c$ is the speed of the light; $n$ is the refraction
index of the transmission medium) \cite{Su,Glo,Wang}. For our experiment,
the measured noise sideband frequency is $f=2$ $MHz$, so $\Delta L$ should
be $48$ $m$. Combining the detected noise powers of the amplitude quadrature
$X_{3}$ and $X_{4}$ ($X_{2}$ and $X_{3}$; $X_{2}$ and $X_{4}$) by means of a negative power combiner (not
shown in the figure) we obtain the correlation variance $\langle
\delta ^{2}(X_{3}-X_{4})\rangle $ ($\langle \delta ^{2}(X_{2}-X_{3}),$ $%
\langle \delta ^{2}(X_{2}-X_{4})\rangle $). Similarly, adding the noise
powers of the phase quadratures with a positive power combiner (not shown in
the figure) the correlation variances $\langle \delta
^{2}(g_{1}^{opt}Y_{2}+Y_{3}+Y_{4})\rangle $ ($\langle \delta
^{2}(Y_{2}+g_{2}^{opt}Y_{3}+Y_{4})\rangle $; $\langle \delta
^{2}(Y_{2}+Y_{3}+g_{3}^{opt}Y_{4})\rangle $) is obtained. At last, the
combined correlations variances of the amplitude and the phase quadratures
are recorded by a spectrum analyzer (not drawn in Fig. 2).

\begin{figure}
\centerline{
\includegraphics[width=86mm]{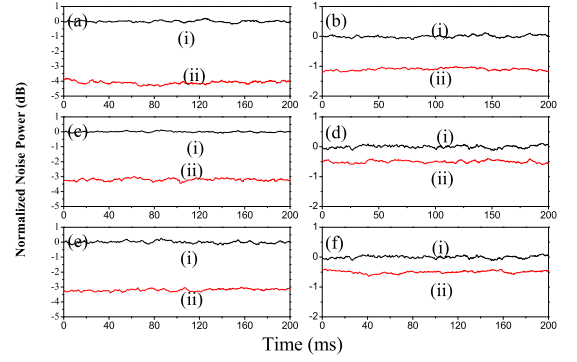}
} \caption{(Color online) The measured correlation variance of
three-color entangled states at 2 $MHz$. \textbf{a}, $\langle \delta ^{2}(X_{3}-X_{4})\rangle $, \textbf{b}, $\langle
\delta ^{2}(g_{1}^{opt}Y_{2}+Y_{3}+Y_{4})\rangle $,
\textbf{c}, $\langle \delta
^{2}(X_{2}-X_{4})\rangle $, \textbf{d},
$\langle \delta
^{2}(Y_{2}+g_{2}^{opt}Y_{3}+Y_{4})\rangle $, \textbf{e}, $\langle \delta
^{2}(X_{2}-X_{3})\rangle $, \textbf{f}, $\langle \delta
^{2}(Y_{2}+Y_{3}+g_{3}^{opt}Y_{4})\rangle $. (i) The QNL; (ii) The
correlation noise power. The resolution and video bandwidths of spectrum analyzer are 30kHz and 100Hz, respectively. \label{Fig3} }
\vspace{-1em}
\end{figure}

During the experiment, we adjusted the temperature of the PPKTP crystal in the NOPO1 to $23.78$ $^0C$ firstly to produce a pair of the signal beam at $746.64$ $nm$ and the idler beam at $852.35$ $nm$, which were split by a beam splitter (BS1). The wavelength of the idler beam was at a transition of the Cs atoms exactly. When the pump power at $398$ $nm$ was $118$ $mW$, the power of the output subharmonic wave was $17$ $mW$. Tuning the temperature of the PPLN in NOPO2 to $154.0$ $^0C$, the wavelength of the signal and the idler beams from the NOPO2 were $1550.60$ $nm$ and $1440.06$ $nm$ respectively, which were split by a beam splitter (BS2). Under the pump power of $14.6$ $mW$, the power of the signal (idler) beam from NOPO2 was $3.2$ $mW$.

 The measured correlation variances of the noise powers of the amplitude and
the phase quadratures among the three resultant optical beams at $852.35$ $nm$, $1550.60$ $nm$ and $1440.06$ $nm$ in term of Eqs. (1) are shown in Fig. 3 (a)-(f), where the traces (i) and
the traces (ii) stand for the QNLs and the correlation variances,
respectively. From Fig. 3 we have $\langle \delta ^{2}(X_{3}-X_{4})\rangle
=-4.1\pm 0.1dB$, $\langle \delta ^{2}(g_{1}^{opt}Y_{2}+Y_{3}+Y_{4})\rangle
=-1.1\pm 0.1dB$, $\langle \delta ^{2}(X_{2}-X_{3})\rangle =-3.2\pm 0.1dB$, $%
\langle \delta ^{2}(Y_{2}+g_{2}^{opt}Y_{3}+Y_{4})\rangle =-0.5\pm 0.1dB$, $%
\langle \delta ^{2}(X_{2}-X_{4})\rangle =-3.2\pm 0.1dB$ and $\langle \delta
^{2}(Y_{2}+Y_{3}+g_{3}^{opt}Y_{4})\rangle =-0.5\pm 0.1dB$. Where, the minus
symbol before the first numbers in the right sides of these equalities means
that the variances are below the corresponding QNL and $g_{1}^{opt}=0.95\pm 0.02$, $g_{2}^{opt}=1.00\pm 0.02$ and $g_{3}^{opt}=1.00\pm 0.02$ stand for the optimal gain factors of $g_{j}$ taken in the experiment for obtaining the highest correlations \cite{Loock,SM}. The three combinations
of the correlation variances in Eqs. (1) are $\Delta _{1}=3.03\pm 0.06$, $\Delta
_{2}=3.68\pm 0.05$ and $\Delta _{3}=3.68\pm 0.05$ respectively and all of them are smaller
than the criterion ''$4$'' for the CV three-partite entanglement of optical
modes. Thus, the three-color CV entanglement is experimentally demonstrated.

For exhibiting the tunable property of NOPO, we measure the function of the
wavelength of the output optical mode from NOPO2 vs the temperature of the
nonlinear crystal (Fig.4). When the temperature of PPLN is changed from $130$
$^{o}C$ to $160$ $^{o}C$ the measured wavelength of the output signal beam ($%
\blacktriangle $) and idler beam ($\blacktriangledown $) are changed from
$1549.0$  $nm$ and $1441.4$ $nm$ to $1550.9$ $nm$ and $1439.8$ $nm$, respectively. The solid
lines are the theoretically fitting under the idea phase matching condition.
The tuned wavelength range is about $1.9$ $nm$ for the signal
beam. If tuning the temperature of the PPKTP inside NOPO1 the similar
tunable property can also be observed. The quality of the three-color
entanglement in the total tuning range is almost the same. The good agreement
between the experimental measurements and the theoretical calculations as
well as the perfect linearity of the dependence of the output wavelength on
the crystal temperature demonstrate that the NOPO is an idea device for
generating the tunable CV entangled states of light.

\begin{figure}
\centerline{
\includegraphics[width=86mm]{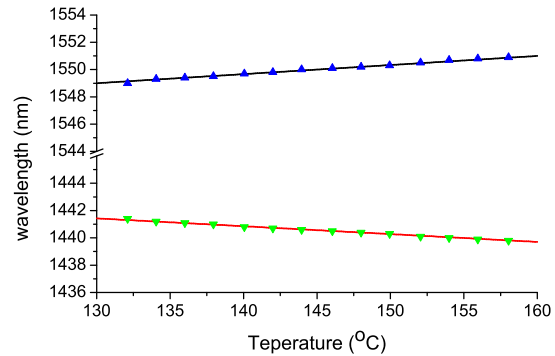}
} \caption{(Color online) Wavelengths of the output signal and idler beams
from the NOPO2 changing with the temperature of the PPLN
crystal. The solid line is the calculated curve; $\blacktriangle$ and $\blacktriangledown$ are the measured wavelengths of the signal and idler beams, respectively. \label{Fig4} }
\vspace{-1em}
\end{figure}

For the conclusion, we achieve the experimental generation of the three-color CV entangled states of optical modes at the Cs atomic transition and the optical fiber communication frequencies. The results numerically calculated in Ref. \cite{Tan} show that the physical parameters of the two NOPOs, such as the transmissions of input and output mirrors, the intra-cavity losses, and the pump coefficient $\sigma$ ($\sigma =\sqrt{P/P_{th}}$, $P$ is the pump power of NOPO, $P_{th}$ is the threshold pump power), simultaneously influence the entanglement quality. For improving entanglement quality the intra-cavity losses of the signal and the idler modes should be reduced. However, the intra-cavity losses of NOPO and the threshold pump power are limited by the quality of the non-linear crystal and the cavity mirrors. Therefore, for a given experimental system the pump powers of two NOPOs are only adjustable parameters. In the experiments we have to compromisingly choose the pump coefficients of the two NOPOs to obtain the possibly highest entanglement (See Supplemental Material \cite{SM} for details). Although the completed entanglement is not very high, the advantage of the wavelengths of the entangled sub-modes is obvious. For example, if we suppose that $-0.2$ $dB$ is a cut-off point of entanglements, the achievable maximum transmission distances of entangled states at $1550$ $nm$, $1440$ $nm$ and $1064$ ($1080$) $nm$ with $-3.2$ $dB$ amplitude correlation in the optical fiber are $52$ $km$, $42$ $km$ and $16$ $km$, respectively (See Supplemental Material \cite{SM} for details). According to the CV quantum key distribution scheme in Ref. \cite{Su2}, the obtainable secure bit rate with $-3.2$ $dB$ amplitude correlation light at $1550$ $nm$ is $3$ $kbits/s$ against the collective attack after passing through a $20$ $km$ optical fiber with the transmission efficiency of $40\%$. However, if the sub-mode of $-3.2$ $dB$ correlation at $1064$ $nm$ is used, the transmission efficiency in the fiber is only $5\%$ and thus no any secure key can be distilled (See Supplemental Material \cite{SM} for details).

Using this system, the entangled states can be prepared under normal phase-matching conditions without the need of cooling dowm the system to below zero degree \cite{Coe}. The free-selectivity of two nonlinear crystals in the two cascaded NOPOs and the tunable property of NOPO provide large ranges of frequency-selection for the three-color entanglement. If NOPOs more than two are cascaded the multi-color CV entangled states with needed frequencies more than three are able to be generated based on the presented method.

This research was supported by National Basic Research Program of China
(Grant No. 2010CB923103), Natural Science Foundation of China (Grants Nos.
11074157, 61121064 and 11174188), the TYAL.

\end{document}